\documentclass[
    preprint,
    aps,
    groupedaddress,
    superscriptaddress,
    amssymb,
    amsmath
]{revtex4-1}

\usepackage{graphicx}   
\usepackage{color}      
\usepackage[
  colorlinks=true,
  urlcolor=blue,
  linkcolor=blue,
  citecolor=blue
]{hyperref}   
\usepackage{siunitx}

\begin{document}
\title{Spin-dependent directional emission from an asymmetry optical waveguide with an embedded quantum dot ensemble}
\author{Wenbo Lin}
\email{lin-w@iis.u-tokyo.ac.jp}
\affiliation{
Institute of Industrial Science, the University of Tokyo, 4-6-1 Komaba, Meguro-ku, Tokyo 153-8505, Japan
}
\author{Yasutomo Ota}
\affiliation{
Institute of Nano Quantum Information Electronics, the University of Tokyo, 4-6-1 Komaba, Meguro-ku, Tokyo 153-8505, Japan
}
\author{Satoshi Iwamoto}
\affiliation{
Institute of Industrial Science, the University of Tokyo, 4-6-1 Komaba, Meguro-ku, Tokyo 153-8505, Japan
}
\affiliation{
Institute of Nano Quantum Information Electronics, the University of Tokyo, 4-6-1 Komaba, Meguro-ku, Tokyo 153-8505, Japan
}
\author{Yasuhiko Arakawa}
\affiliation{
Institute of Nano Quantum Information Electronics, the University of Tokyo, 4-6-1 Komaba, Meguro-ku, Tokyo 153-8505, Japan
}
\date{\today}
\begin{abstract}
In this study, we examine a photonic wire waveguide embedded with an ensemble of quantum dots that directionally emits into the waveguide depending on the spin state of the ensemble. This is accomplished through the aid of the spin-orbit interaction of light. The waveguide has a two-step stair-like cross section and embeds quantum dots (QDs) only in the upper step, such that the circular polarization of emission from the spin-polarized QDs controls the direction of the radiation. We numerically verify that more than 70\% of the radiation from the ensemble emitter is toward a specific direction in the waveguide. We also examine a microdisk resonator with a stair-like edge, that supports selective coupling of the QD ensemble radiation into a whispering galley mode rotating unidirectionally. Our study provides a foundation for spin-dependent optoelectronic devices.
\end{abstract}
\pacs{xx.xx+x}
\maketitle
The spin-orbit interaction (SOI) of light mediates the spin-to-orbital conversion of the angular momentum of light, which is similar to its well-known electronic counterpart~\cite{bliokh2015spin}. Strong optical SOIs emerge in optical fields with large intensity gradients, by which the spin and orbital angular momentum of light couple each other and become inseparable quantities. The coupling of the angular momentum has been observed in various systems such as tightly focused optical beams~\cite{rodriguez2010optical} and optical modes confined in photonic nanostructures~\cite{petersen2014chiral,o2014spin,fong18scheme}. As a result of the coupling, the spin angular momentum or the circular polarization of light, can control the angular or linear momentum of the optical modes excited in the system. This SOI-based functionality has been employed to demonstrate dielectric and plasmonic optical mode converters~\cite{marrucci2006optical,karimi2014generating} and beam separators~\cite{lin2013polarization} that are regulated by the circular polarization of incident light. Furthermore, optical SOI can also be exploited to control photon emission and absorption processes, particularly when the emitter supports a well-isolated spin state and emits circularly polarized photons. It has been demonstrated that this type of single quantum emitter radiates unidirectionally into a waveguide~\cite{mitsch2014quantum,luxmoore2013interfacing,sollner2015deterministic,coles2016chirality} when positioned at a chiral point, or a \textit{C}-point~\cite{young2015polarization,coles2016chirality} in the waveguide, which derives from the optical SOI.

Extension of the optical SOI effect to ensemble emitters is a fascinating research direction, as it may realize various novel optoelectronic devices, such as spin-controllable unidirectional lasers and amplifiers. The optical functionalities regulated through the electron spin degree of freedom could be of importance in, for example, future on-chip optical interconnection in spintronics. Recently, the spin-dependent control of spontaneous/stimulated emission processes in ensemble emitters has been demonstrated using photonic and plasmonic structures~\cite{mitsch2014quantum,takahashi2017circularly,zambon2018optically,gong2018nanoscale,spitzer2018routing}. For application to integrated photonics, implementing this type of radiation control is desirable in photonic nano/microstructures that are compatible with planar optical circuits and are composed of low-loss dielectric materials. For practical use, the ease of the fabrication of the structures is also important.

In this study, we theoretically examine a photonic waveguide structure embedding an ensemble of spin-polarized quantum dots (QDs), which undergo directional spontaneous emission into the waveguide. Considering practical applications, we examine devices embedded with epitaxially grown QDs as representative materials that fit with photonic integration and faithfully convert electronic spin polarization into photon circular polarization~\cite{lodahl2015interfacing,paillard2001spin,braun2005direct}. The waveguide has an asymmetric cross section that resembles two-step stairs. We consider the case in which the QDs are distributed only within the upper step. This symmetry breaking in the QD distribution provides directionality to the spontaneous emission from the QD ensemble. Through numerical simulations, we show that more than 70\% of the total spontaneous emission is toward a particular propagation direction of the waveguide while avoiding a significant reduction of the total spontaneous emission rate of the ensemble emitter. We also apply the stair-shaped structure in building a microdisk resonator. We observe that the spin-polarized QD ensemble enclosed in the resonator selectively couples to a whispering galley mode that rotates in a particular direction, which may enable spin-controllable non-reciprocal gain and lasing.

\begin{figure}[tb]
\centering
\includegraphics[width=\linewidth]{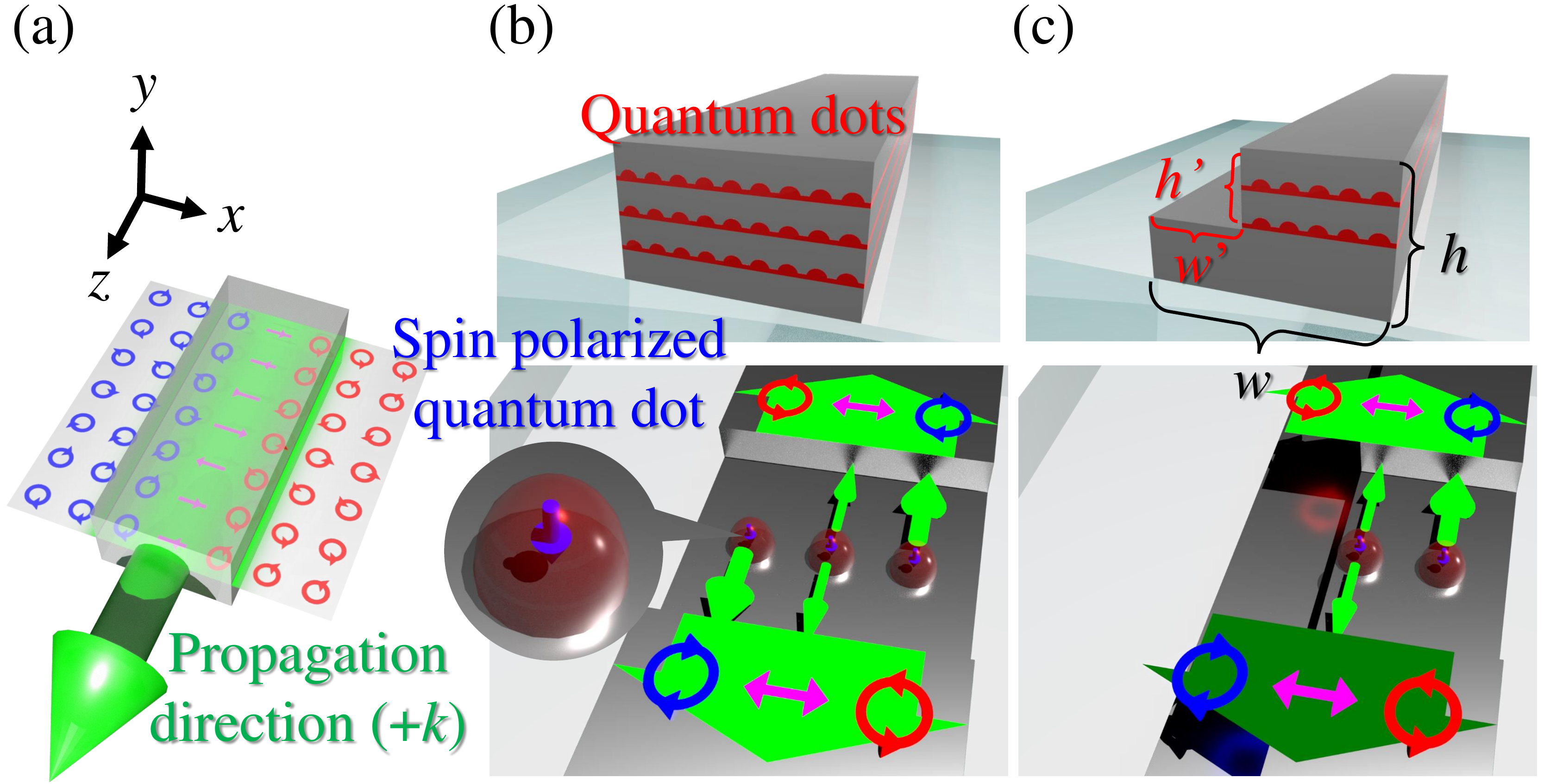}
\caption{%
\textbf{(a)}~Polarization map for the fundamental TE-like mode propagating in the $+z$ direction.
\textbf{(b)}~Schematic cross-section of a rectangular wire waveguide embedding QDs. The bottom inset schematically shows the manner in which the radiation from circularly polarized QDs directs within the waveguide.
\textbf{(c)}~Schematic cross section of the waveguide with a two-step stair-like cross section that embeds QDs in the upper step. The asymmetric QD distribution leads to directional radiation from the spin-polarized ensemble QDs (bottom inset).%
}
\hrule height.5pt
\label{fig:concept}
\end{figure}

We initiate our study using a conventional wire waveguide that supports a confined optical mode with transverse-electric (TE)-like polarization. Because of the spatial confinement and the resulting optical SOI, the electric field at each location in the waveguide evolves with a polarization that differs from that of plane waves. Figure~\ref{fig:concept}(a) show a schematic of the polarization distribution of a waveguide mode that propagates in a particular direction. Linear polarization is found at the center of the waveguide, whereas circular polarization resides near the waveguide edge. The handedness of the circular polarization flips across the waveguide center: right-handed $\left( \sigma^+ \right)$ and left-handed $\left( \sigma^- \right)$ circular polarizations are found near the right left edges, respectively. The circular polarization in the waveguide constitutes a transverse spin~\cite{aiello2009transverse,bliokh2015transverse} of light, the rotation axis of which is perpendicular to the propagation direction. The locations supporting the pure circular polarization constitute \textit{C}-points. For the same waveguide mode traveling to the opposite direction, the handedness of the circular polarization inverts anywhere in its field distribution, as a result of time-reversal symmetry. We then consider spontaneous emission from the QDs uniformly embedded in the rectangular waveguide, a schematic of which is shown in Fig.~\ref{fig:concept}(b). We assume that the QDs can be expressed by TE-polarized optical transition dipoles, which is typical for epitaxially grown QDs by the Stranski-Krastanov mode~\cite{lodahl2015interfacing}. When the QDs are homogeneously spin-polarized, each QD behaves as a circular dipole oscillating in the same direction in the plane where the QDs distribute. In this case, those QDs located near the waveguide center couple equally to both the waveguide modes propagating in opposite directions. Those QDs near one of the waveguide edges directionally radiate into a propagating mode. However, the remaining QDs located near the other side couple to the opposite directing propagating mode, resulting in a vanishing directionality as a whole. Consequently, a specific type of symmetry breaking is necessary to achieve directional light emission.

To obtain directional radiation from the QD ensemble, we propose using a waveguide structure with an asymmetric two-step stair-like cross section as shown in Fig.~\ref{fig:concept}(c). The waveguide has an upper step with a depth of $h'$ and a width of $w-w'$, in which QDs are assumed to distribute uniformly. The asymmetric distribution of the QDs enables directional radiation under spin-polarized excitation of the QDs. It is important to note that, this structure can be easily fabricated using standard lithography processes consisting of dry etching into conventional epitaxially grown QD wafers, which can avoid reliance on complicated processing such as crystal regrowth~\cite{matsuo2010high,vo2018two}. Hereafter, we discuss directional emission from the QDs in the waveguide structure using numerical simulations.

We evaluate the directionality of the QD emission using the two quantities defined as follows. The first is the direction-resolved spontaneous emission rate of a spin-polarized QD, $\gamma_\pm$. Here, the suffix $+(-)$ expresses the mode propagating into a $+k(-k)$ direction, to which the QD emission couples. $\gamma_\pm$ for a QD located at $\boldsymbol{r}$ is given by~\cite{verhart2014single}:
\begin{equation}
\gamma_{\pm}\left( \boldsymbol{r} \right) = \frac{\omega}{4\hbar c}n_{g}\frac{\left| \boldsymbol{\mu}^\ast \cdot \boldsymbol{E}_{\pm}\left( \boldsymbol{r} \right) \right|^2}{\iint \varepsilon_0 \varepsilon\left( \boldsymbol{r} \right) \left| \boldsymbol{E}_{\pm}\left( \boldsymbol{r} \right) \right|^2 \mathrm{d}\boldsymbol{S}},
\label{eq:sponrate}
\end{equation}
where $\omega$ is the angular frequency of radiation, $\hbar$ is the reduced Planck's constant, $c$ is the velocity of light, $n_g$ is the group velocity of the waveguide mode, $\boldsymbol{\mu}$ is the optical transition dipole moment with certain circular polarization, $\boldsymbol{E}_{\pm}\left( \boldsymbol{r} \right)$ is the electric field of the waveguide mode propagating to the $\pm k$ direction, $\varepsilon_0$ is the permittivity of the vacuum, and $\varepsilon\left( \boldsymbol{r} \right)$ is the relative permittivity at the location of the QD. The integration in the denominator is conducted for the entire waveguide cross section. At a \textit{C}-point, $\gamma_\pm$ for a spin-polarized QD is maximized or minimized depending on the handedness of the QD circular dipole. We also consider the total spontaneous emission rate of the QD ensemble, $\Gamma_\pm$, which is defined by the integration of Eq.~(\ref{eq:sponrate}) for all QDs in the waveguide cross section. As the second quantity, we introduce a degree of unidirectionality (DOU), which is defined as:
\begin{equation}
\mathrm{DOU} = \frac{\gamma_{+} - \gamma_{-}}{\gamma_{+} + \gamma_{-}}.
\label{eq:defDOU}
\end{equation}
This indicates the fraction of the directional radiation to the radiation of a QD into the waveguide. By replacing $\gamma_\pm$ with $\Gamma_\pm$ in the equation, a DOU can measure the directionality of emission from a QD ensemble.

\begin{figure}[tb]
\centering
\includegraphics[width=\linewidth]{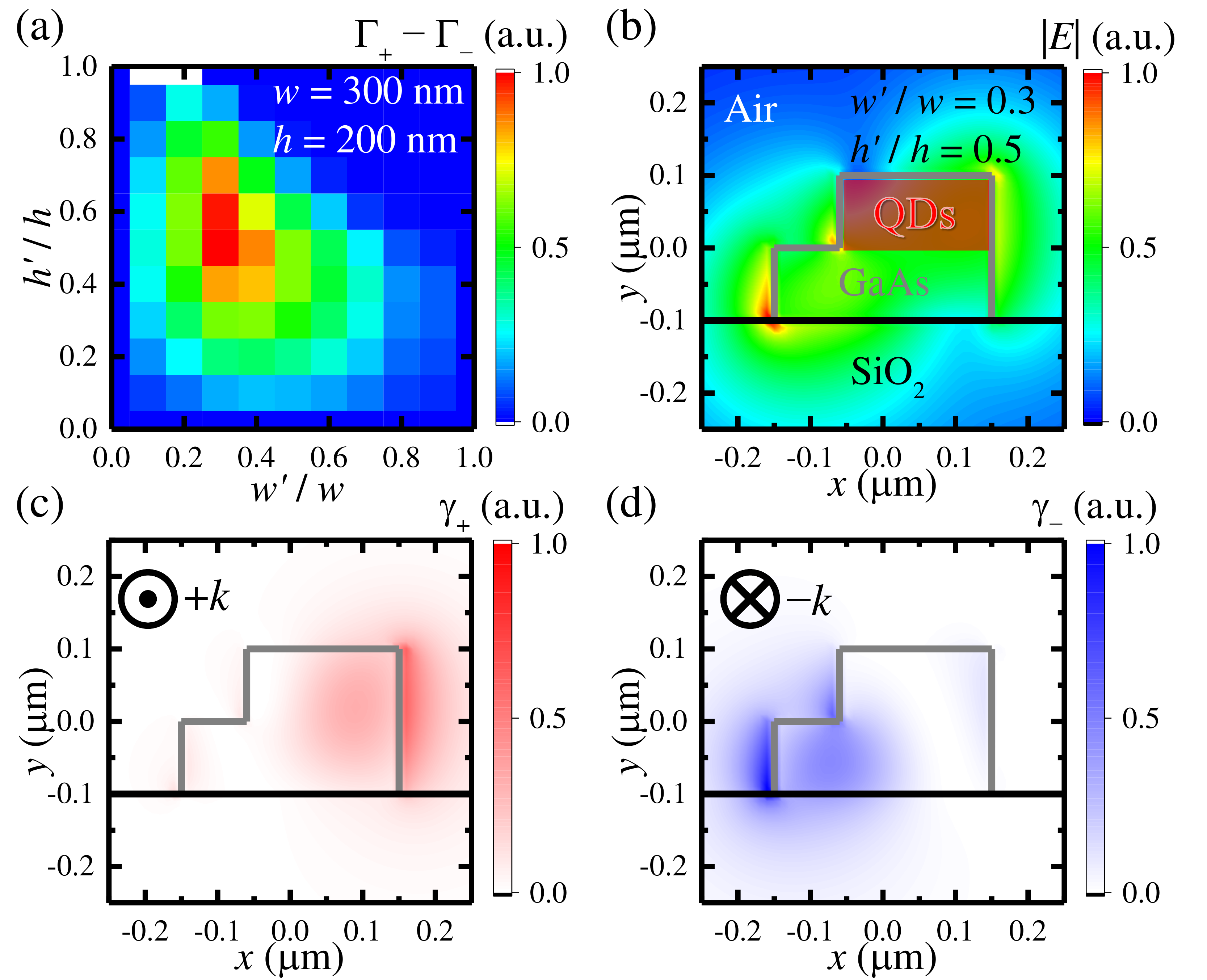}
\caption{%
\textbf{(a)}~Computed differences of the direction-resolved spontaneous emission rates of the QD ensemble, $\Gamma_{+} - \Gamma_{-}$.
\textbf{(b)}~Simulated mode profile for the waveguide designed with $w'$~=~90~\si{\nano\meter} and $h'$~=~100~\si{\nano\meter}.
\textbf{(c)--(d)}~Maps of the direction-resolved spontaneous emission rates, $\left(\gamma_{\pm}\right)$, calculated for a $\sigma^+$ QD coupled to (c) the mode propagating the $+k$ direction and to (d)~the $-k$ direction.
The refractive indices of GaAs and glass are assumed to be 3.4 and 1.5, respectively.%
}
\hrule height.5pt
\label{fig:fem}
\end{figure}
Our design starts with a symmetric GaAs waveguide on glass with a rectangular cross section having a width of $w = 300$~\si{\nano\meter} and height of $h = 200$~\si{\nano\meter}, which supports a single optical mode at a wavelength of 1300~\si{\nano\meter}. The upper region of the waveguide is assumed to enclose QDs uniformly. Because of the inversion symmetry of the waveguide system with respect to the propagation axis, the QD ensemble in this waveguide does not exhibit directional emission. We induce directionality in the QD emission by etching out the $w' \times h'$ region, as shown in Fig.~\ref{fig:concept}(c). The remaining upper region $\left( w - w' \right) \times h'$ encloses the QDs. With varying $w'$ and $h'$, we first sought the optimal design for direction emission. For each parameter set, we computed the waveguide mode profile at the cross section using the finite element method (FEM) and then calculated the spontaneous emission rate of the QD ensemble. Figure~\ref{fig:fem}(a) shows the computed differences of the direction-resolved spontaneous emission rates of the QD ensemble, $\Gamma_{+} - \Gamma_{-}$. Each QD was assumed to radiate as a $\sigma^+$ dipole at 1300~\si{\nano\meter}. We observed that the difference grows as the etching region increased to approximately $\left( w'/w, h'/h \right) = \left( 0.5, 0.5 \right)$. Further enlarging the etching region reduced the total number of the QDs in the structure considerably, which in turn diminished the total spontaneous emission rate and reduce the magnitude of the difference of the emission rates. The difference was maximized at $\left( w'/w, h'/h \right) = \left( 0.3, 0.5 \right)$, where the ratio $\Gamma_+/\Gamma_-$ reached 3. For this parameter set, we computed a mode profile as shown in Fig.~\ref{fig:fem}(b). A considerable mode overlap with the QD ensemble is seen. We also simulated the maps of $\gamma_+$ and $\gamma_-$ for $\sigma^+$ dipoles located in the waveguide mode, as plotted in Figs~\ref{fig:fem}(c) and (d). The region embedding the QDs was well within the area where $\gamma_+$ dominates. In addition, the area with pronounced $\gamma-$ located predominantly around the lower step. The contrast between the $\gamma_+$ and $\gamma_-$ distributions induced a large $\Gamma_+ - \Gamma_-$ in this design. The total emission rate to a particular direction $\left( \Gamma_+ \right)$ was comparable ($\approx$74\%) with  that of the unetched rectangular waveguide that embeds the QDs in the upper region. We note that in this parameter set, a transverse-magnetic (TM)-like mode was guided in the structure, but its coupling to the QDs was marginal because of the mismatch between the QD's dipole orientation and the electric field polarization of the mode.

\begin{figure}[b]
\centering
\includegraphics[width=\linewidth]{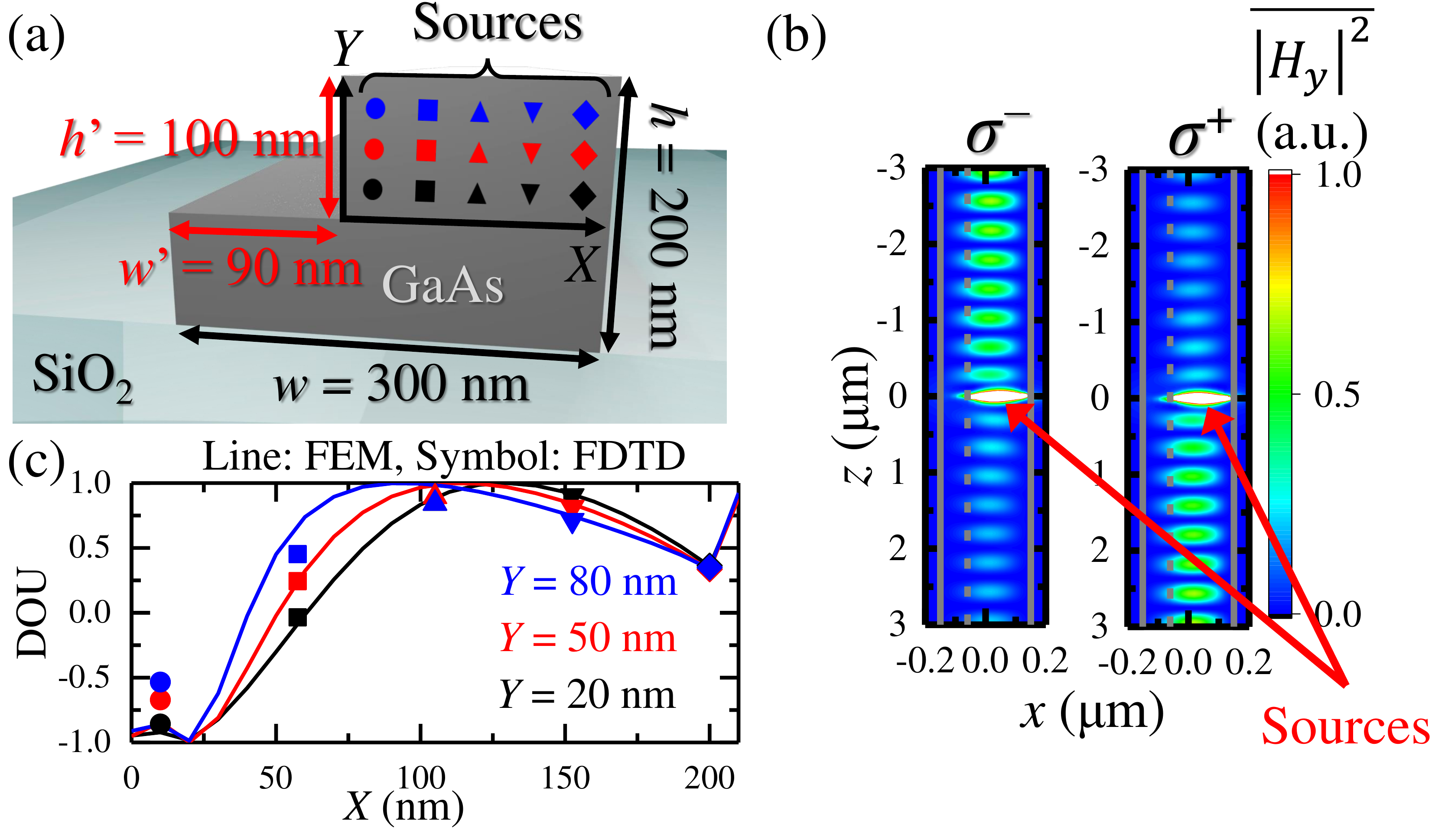}
\caption{%
\textbf{(a)}~FDTD model employed for the full 3D simulations. The positions of the dipole sources are indicated by the colored symbols. The coordinate axes are also indicated.
\textbf{(b)}~Averaged field patterns of $\left| H_y \right|^2$ calculated using the 15 circular dipoles with $\sigma^-$ (left panel) or $\sigma^+$ (right panel) polarization. The field slices are taken at 100~\si{\nano\meter} above the glass-GaAs interface.
\textbf{(c)}~Position-resolved DOU values computed by the FDTD method (symbols). For comparison, the corresponding FEM simulation results are indicated by the solid lines.%
}
\label{fig:fdtd}
\end{figure}
We further verified these results by performing full three-dimensional simulations of spontaneous emission based on the finite-difference time-domain (FDTD) method~\cite{vuckovic1999finite}. The parameters of the waveguide were the same as those that maximize $\Gamma_+ - \Gamma_-$ in the FEM calculations. We emulated the QD ensemble with three layers of the array of circular dipoles as depicted in Fig.~\ref{fig:fdtd}(a). The layer distance was set to 30~\si{\nano\meter}, which was easily realized using conventional crystal growth methods. We avoided unwanted interference between the coherent circular dipoles in the simulator by calculating the contribution of each dipole one by one and later merging them numerically. The two panels in Fig.~\ref{fig:fdtd}(b) show the magnetic field profiles computed for the 15 QDs with $\sigma^-$ and $\sigma^+$ polarization, respectively. Directional emission from the ensemble that depends on the circular polarization can be clearly seen. Measured DOU of this case was 0.42, which roughly agrees with the prediction by the FEM simulation (i.e., 0.59). A DOU of 0.42 means that 71\% of QD emission coupled to the waveguide is toward a particular direction. Figure~\ref{fig:fdtd}(c) plots the calculated DOU of each circular dipole distributed in the QD layers. For comparison, we also plotted the evolution of DOU along with the QD layers computed by the FEM. The two plots are comparable, but some deviations can be found, particularly around $X \approx 0$. These discrepancies chiefly originated from the coupling to a TM-like mode, which has been neglected in previous FEM analysis. Indeed, wuth the contribution from the TM mode, the results of the two types of simulations coincide remarkably (not shown). Other possible sources of the deviation between the FEM and FDTD simulations are the coupling of the QDs to other leaky modes and the sparse arrangement of the QDs in the FDTD simulator.

\begin{figure}[tb]
\centering
\includegraphics[width=\linewidth]{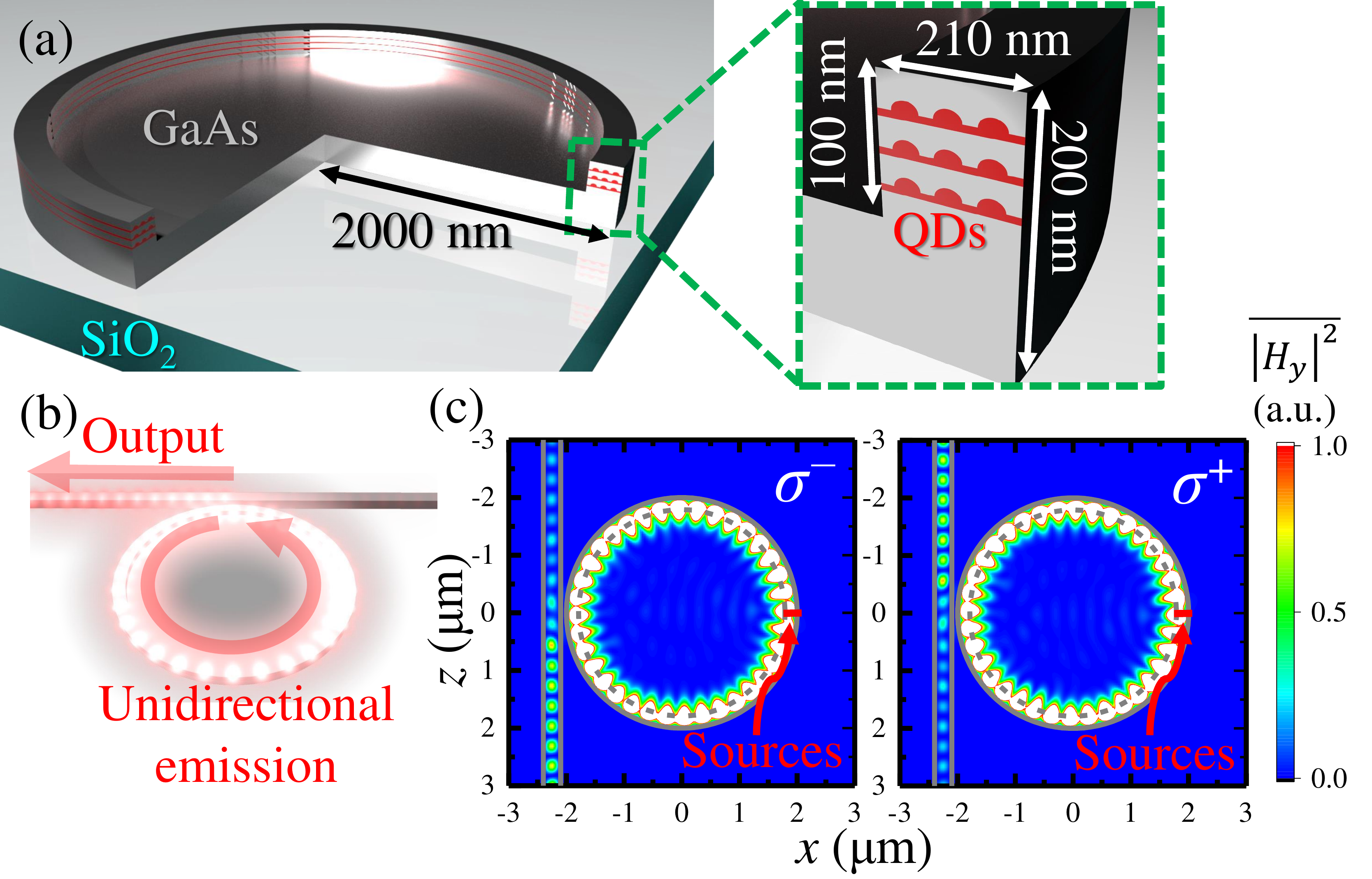}
\caption{%
\textbf{(a)}~Schematic of the disk resonator with a stair-shaped edge. The inset shows a close-up of around the edge.
\textbf{(b)}~Schematic of the disk resonator coupled to an output waveguide.
\textbf{(c)}~Simulated coupling of the QD ensemble to the WGM modes, which plot the averaged field patterns of $\left| H_y \right|^2$ calculated from the 15 circular dipoles distributed in the cross section with $\sigma^-$ (left panel) or $\sigma^+$ (right panel) polarization.%
}
\hrule height.5pt
\label{fig:disk}
\end{figure}
Next, we discuss optical resonators based on the asymmetric waveguide that embeds an ensemble of QDs. We investigated a whispering galley mode (WGM) disk resonator on glass as schematically shown in Fig.~\ref{fig:disk}(a). The edge of the disk has a step which encloses an ensemble of spin-polarized QDs. We considered a disk with a diameter of 2~\si{\micro\meter}, and a thickness of 200~\si{\nano\meter}. The height and width of the step were 100 and 210~\si{\nano\meter}, respectively, which coincide with the dimensions of the step discussed in Fig.~\ref{fig:fdtd}. We considered that the biased distribution of the QDs enables selective coupling of QD spontaneous emission into one of the degenerated WGMs rotating either clockwise or counter-clockwise. The disk resonator evanescently couples to a wire waveguide across a 100-\si{\nano\meter} air gap, which was employed to highlight the directional output from the resonator, as shown in Fig.~\ref{fig:disk}(b). Without the waveguide, the disk resonator supports a WGM of an azimuthal order of 18 with a high $Q$ factor of approximately $3.3 \times 10^4$ and a large free spectral range of $\approx 80$~\si{\nano\meter} around a wavelength of 1.31~\si{\micro\meter}. With coupling to the waveguide, the same WGM reduces the $Q$ factor to $1.6 \times 10^3$. Based on this design, we investigated selective coupling of QD radiation to the WGMs using the FDTD method. We followed the same procedure employed for calculating the plots in Fig.~\ref{fig:fdtd}(b). We considered a set of QD excitation sources in a cross section of the disk arranged in the same manner as previously described. We assumed that all QDs radiated resonantly to the WGMs at 1.31~\si{\micro\meter} and then calculated steady state field distributions, the average of which is shown in Fig.~\ref{fig:disk}(c). Under the $\sigma^+$ dipole excitation, the clockwise WGM was selectively excited, resulting in an upstream output into the adjacent waveguide. When we switched to the $\sigma^-$ excitation, the reverse rotating mode was excited as expected. The obtained DOU was 0.51, which was higher than that measured for the straight waveguide. This improvement was likely due to the decoupling of relevant TM-like WGMs from the resonance of the TE-like WGMs under investigation. The DOU of 0.5 corresponded to the ratio $\Gamma_+ / \Gamma_-$ of 3 under the $\sigma^+$ excitation. This large imbalance of the total spontaneous emission rates of the QD ensemble is equal to the selective gain supply to the clockwise mode, which provides a basis for realizing unidirectional lasers and amplifiers. In actual lasing devices, in addition to the unequal mode gains, the cross-gain saturation as observed in other spin lasers~\cite{ando1998photon,zambon2018optically} may facilitate a unidirectional operation since it further suppresses the growth of the other mode. Indeed, given the simulation results and a simple set of two-mode laser equations, we confirmed that $\Gamma_+ / \Gamma_- = 3$ is sufficient to attain a strong unidirectionality when lasing (not shown).

In summary, we examined an asymmetric waveguide that enables directional spontaneous emission of QDs embedded in the structure. The waveguide has a cross section resembling two-step stairs and facilitates directional coupling from the spin-polarized QD ensemble with the aid of optical SOI. We identified a waveguide design that maximizes $\Gamma_+ - \Gamma_-$ based on the FEM simulations. With this design and when using the FDTD method, we numerically obtained a large DOU of 0.42, which corresponds that 71\% of the spontaneous emission coupled to a particular direction of the waveguide mode. We then applied the asymmetric waveguide to design a disk resonator and observed selective coupling of enclosed QDs to a specific WGM. In this resonator, the ratio of $\Gamma_+ / \Gamma_-$ reached $\approx 3$, thus providing a means of realizing spin-dependent directional lasers and amplifiers. We believe that such spin-dependent unidirectional lasers will be of importance for diverse applications, including optical and magnetic field sensing and future optical interconnections in spintronics.

\bigskip
\noindent\textbf{Funding.} This work was supported by JSPS KAKENHI Grant-in-Aid for Specially promoted Research (15H05700), KAKENHI (15H05868, 16K06294) and the New Energy and Industrial Technology Development Organization (NEDO) project.

\bibliography{OptLett_1_ref}

\end{document}